\documentclass[12pt]{article}
\usepackage{amsmath}
\usepackage{amssymb}
\usepackage{amsfonts}
\usepackage{amscd}
\usepackage{graphics}

\newcommand{\field}[1]{\mathbb{#1}}
\newcommand{\N}{\field{N}}
\newcommand{\R}{\field{R}}

\title{Infinite families of position-dependent mass Schr\"odinger equations with known ground and first excited 
states}

\author{C.\ Quesne\thanks{E-mail address: cquesne@ulb.ac.be} \\
{\small\sl Physique Nucl\'eaire Th\'eorique et Physique Math\'ematique, 
Universit\'e Libre de Bruxelles,} \\ 
{\small \sl Campus de la Plaine CP229, Boulevard~du Triomphe, B-1050
Brussels, Belgium}}
\date{ }
\begin{document}
\baselineskip=22pt plus 1pt minus 1pt
\maketitle

\begin{abstract}
A construction method of infinite families of quasi-exactly solvable position-dependent mass Schr\"odinger equations with known ground and first excited states is proposed in a deformed supersymmetric background. Such families correspond to extensions of known potentials endowed with a deformed shape invariance property. Two different approaches are combined. The first one is a generating function method, which enables to construct the first two superpotentials of a deformed supersymmetric hierarchy, as well as the first two partner potentials and the first two eigenstates of the first potential, from some generating function $W_+(x)$ [and its accompanying function $W_-(x)$]. The second approach is the conditionally deformed shape invariance method, wherein the deformed shape invariance property of the starting potentials is generalized to their extensions by adding some constraints on the parameters and by imposing compatibility conditions between sets of constraints. Detailed results are given for some extensions of the linear and radial harmonic oscillators, as well as the Kepler-Coulomb and Morse potentials.
\end{abstract}

\vspace{0.5cm}

 
\newpage
%
%
\section{Introduction}

The concept of position-dependent mass (PDM) is known to play an essential role in many physical problems, such as the study of electronic properties of semiconductor heterostructures \cite{bastard, weisbuch}, quantum wells and quantum dots \cite{serra, harrison}, helium clusters \cite{barranco}, graded crystals \cite{geller}, quantum liquids \cite{arias}, metal clusters \cite{puente}, nuclei \cite{ring, bonatsos}, nanowire structures \cite{willatzen}, and neutron stars \cite{chamel}.\par
%
%
{}Furthermore, PDM presence in quantum mechanical problems may also reflect some other unconventional effects, such as a deformation of the canonical commutation relations or a curvature of the underlying space \cite{cq04}. It may also appear in the Hermitian Hamiltonian equivalent to a $PT$-symmetric one \cite{jones, mosta}.\par
%
%
All these developments  have stimulated the search for exact solutions of PDM Schr\"odinger equations because they may provide a conceptual understanding of some physical phenomena, as well as a testing ground for some approximation schemes. The generation of PDM and potential pairs leading to such exact solutions has been achieved by extending the methods known in the constant mass case (see, e.g., \cite{cq06} and references quoted therein).\par
%
%
Let us recall that, in the latter case, one distinguishes between exactly solvable (ES), quasi-exactly solvable (QES) and conditionally exactly solvable (CES) Schr\"odinger equations. For ES potentials, all the eigenstates can be found explicitly by algebraic means. In the QES case, for some ad hoc couplings only a finite number of eigenstates can be derived in this way, while the remaining ones can only be obtained through numerical calculations. CES potentials lie somewhere in between, because all their energy levels can be exactly obtained, as with the ES potentials, but their parameters cannot be arbitrarily chosen, as with the QES ones.\par
%
%
Considering more specifically QES potentials, several methods have been used for their construction, among which one may mention an sl(2,$\R$) algebraic approach (valid for the simplest cases connected with polynomial solutions of the Heun equation) \cite{turbiner}, the functional Bethe ansatz method (also working for more complicated cases related to generalizations of this equation) \cite{zhang}, the supersymmetric (SUSY) method \cite{cooper} for generating some QES potential from a known one \cite{gango}, the conditional shape invariance (CSI) symmetry method \cite{chakra, bera} generalizing the shape invariance (SI) concept \cite{genden} known in SUSY, or the use of some generating function in a SUSY background \cite{tkachuk}.\par
%
%
In the PDM case to be considered in the present paper, we will use an extension of the methods employed in Refs.~\cite{chakra, bera} and \cite{tkachuk}. Due to the known equivalence of PDM problems to those arising from a deformation of the canonical commutation relations~\cite{cq04}, the SUSY approach of such methods is changed into a deformed supersymmetric (DSUSY) one and the SI concept, whenever applicable, becomes a deformed shape invariance (DSI) one. Such a framework was previously proposed for generating several ES potentials associated with a PDM backbround \cite{bagchi}.\par
%
%
We plan to construct here infinite families of QES extensions of some of these ES potentials with known ground and first excited states. For such a purpose, we will show that, for a given PDM, the first two superpotentials  of a DSUSY hierarchy can be built from some generating function $W_+(x)$ [and its accompanying function $W_-(x)$], as is the case for the corresponding problem arising in the constant mass case 
\cite{tkachuk}.\footnote{After completion of the present work, Prof.\ V.M.\ Tkachuk drew the attention of the author to the fact that the approach of Ref.~\cite{tkachuk} had already been extended to the case of a PDM presence in Ref.~\cite{voznyak}.} Guessing a pair $(W_+(x), W_-(x))$ appropriate for an infinite family of potential extensions can be achieved by studying in detail the first two members of such an extension family by means of the conditionally deformed shape invariance (CDSI) method, which generalizes the procedure known for constant mass problems \cite{chakra, bera}. Such a method is based upon the observation that the first two members of the family may be endowed with a DSI property provided some constraint conditions relating the potential parameters are satisfied. The compatibility conditions between the two sets of constraint conditions, corresponding to the first two steps of the DSUSY hierarchy, then lead to potentials with known ground and first excited states. From the associated functions $W_+(x)$ and $W_-(x)$, we plan to show that the form of such functions valid for any member of the family can be inferred, which will enable us to solve the problem in full generality.\par
%
%
This paper is organized as follows. In Section 2, the description of PDM Schr\"odinger equations in DSUSY and the DSI property are reviewed. In Section 3, the generating function method for constructing PDM Schr\"odinger equations with known ground and first excited states is presented. In Section 4, the CDSI approach for solving the same problem is explained and combined with the previous method by considering an example of extension family. Some other extension families are then treated by the same procedure in Section 5. Finally, Section 6 contains the conclusion.\par
%
%
\section{Position-dependent mass Schr\"odinger equations and deformed supersymmetry} 

In one-dimensional nonrelativistic quantum mechanics, one may deform the conventional commutation relation $[\hat{x}, \hat{p}] = {\rm i}$, where $\hbar=1$ and $\hat{x}$, $\hat{p}$ may be represented by $x$, $-{\rm i} d/dx$, respectively, into \cite{cq04}
\begin{equation}
  [\hat{x}, \hat{\pi}] = {\rm i} f(x).
\end{equation}
Here $f(x)$ is some positive and smooth parameter-dependent deformation function and $\hat{x}$, $\hat{\pi}$ are assumed to be Hermitian operators with respect to the measure $dx$. They may be represented by $x$ and $- {\rm i} \sqrt{f(x)} (d/dx) \sqrt{f(x)}$, respectively.\par
%
%
By substituting $\hat{\pi}^2$ for $\hat{p}^2$ in the conventional Schr\"odinger equation, we arrive at a deformed equation
\begin{align}
  (\hat{H} - E) \psi(x) &= (\hat{\pi}^2 + V(x) - E) \psi(x) \nonumber \\ 
  &= \left(- \sqrt{f(x)} \frac{d}{dx} f(x) \frac{d}{dx} \sqrt{f(x)} + V(x) - E\right) \psi(x) = 0. \label{eq:def-SE}
\end{align}
Such an equation may also be considered as a PDM Schr\"odinger equation
\begin{equation}
  \left(- m^{-1/4}(x) \frac{d}{dx} m^{-1/2}(x) \frac{d}{dx} m^{-1/4}(x) + V(x) - E\right) \psi(x) = 0,
  \label{eq:PDM-SE}
\end{equation}
where the ordering of the mass $m(x) = 1/f^2(x)$ and the differential operator $d/dx$ is that chosen by Mustafa and Mazharimousavi \cite{mustafa}. As well known, other orderings are possible \cite{vonroos} and can be taken care of by replacing $V(x)$ by $V_{\rm eff}(x)$. For instance, for the so-called BenDaniel and Duke ordering \cite{bendaniel}, for which some physical arguments have been put forward \cite{levy}, Eq.~(\ref{eq:PDM-SE}) is rewritten as
\begin{equation}
  \left(- \frac{d}{dx} \frac{1}{m(x)} \frac{d}{dx} + V_{\rm eff}(x) - E\right) \psi(x) = 0, \qquad V_{\rm eff}(x)
  = V(x) - \frac{1}{2} f(x) f''(x) - \frac{1}{4} f^{\prime 2}(x).
\end{equation}
\par
%
%
In the present paper, our starting point will be Eq.~(\ref{eq:def-SE}) (or the equivalent Eq.~(\ref{eq:PDM-SE})), where $x$ varies in some interval $(x_1,x_2)$, which depends on the choice of mass and of potential. As for the conventional Schr\"odinger equation, bound state wavefunctions $\psi_n(x)$ must be square integrable on the interval of definition $(x_1,x_2)$ with respect to the measure $dx$. However, in addition, they must also ensure the Hermiticity of $\hat{H}$ or, equivalently, that of $\hat{\pi}$, which imposes that \cite{bagchi}
\begin{equation}
  |\psi_n(x)|^2 f(x) = \frac{|\psi_n(x)|^2}{\sqrt{m(x)}} \to 0 \qquad \text{for $x \to x_1$ and $x \to x_2$.}
  \label{eq:Hermiticity}
\end{equation}
This represents an additional restriction whenever $f(x) \to \infty$ (or $m(x) \to 0$) for $x \to x_1$ and/or $x \to x_2$.\par
%
%
Equation (\ref{eq:def-SE}) can be discussed in terms of DSUSY \cite{bagchi}. In the simplest case of unbroken DSUSY, one introduces a rescaled potential
\begin{equation}
  V_1(x) = V(x) - E_0,  \label{eq:V1}
\end{equation}
where $E_0$ is the ground state energy of (\ref{eq:def-SE}), and one considers a pair of partner Hamiltonians
\begin{equation}
  \hat{H}_{1,2} = \hat{\pi}^2 + V_{1,2}(x) + E_0, \qquad V_{1,2}(x) = W^2(x) \mp f(x) \frac{dW}{dx},
  \label{eq:V1,2}
\end{equation}
defined on the same interval $(x_1,x_2)$. The superpotential $W(x)$ is related to the ground state wavefunction $\psi_0(x)$ of $\hat{H}_1$ through
\begin{equation}
  W(x) = - f(x) \frac{d}{dx} \log \psi_0(x) - \frac{1}{2} \frac{df}{dx}
\end{equation}
or, conversely,
\begin{equation}
  \psi_0(x) \propto f^{-1/2} \exp\left(- \int^x \frac{W(x')}{f(x')} dx'\right).  \label{eq:psi0}
\end{equation}
\par
%
%
The two partner Hamiltonians can be written as
\begin{equation}
  \hat{H}_1 = \hat{A}^+ \hat{A}^- + E_0, \qquad  \hat{H}_2 = \hat{A}^- \hat{A}^+ + E_0,
\end{equation}
in terms of a pair of first-order differential operators
\begin{equation}
  \hat{A}^{\pm} = \mp \sqrt{f(x)} \frac{d}{dx} \sqrt{f(x)} + W(x),  \label{eq:A}
\end{equation}
and they intertwine with $\hat{A}^+$ and $\hat{A}^-$ as
\begin{equation}
  \hat{A}^- \hat{H}_1 = \hat{H}_2 \hat{A}^-, \qquad \hat{A}^+ \hat{H}_2 = \hat{H}_1 \hat{A}^+.
\end{equation}
The operator $\hat{A}^-$ annihilates the ground state wavefunction $\psi_0(x)$ of $\hat{H}_1$, whereas $\hat{A}^+$ transforms the ground state wavefunction $\psi'_0(x)$ of $\hat{H}_2$ into the first excited state wavefunction $\psi_1(x)$ of $\hat{H}_1$.\par
%
%
On iterating the procedure by considering $\hat{H}_2$ as the new starting Hamiltonian, one may in principle obtain another DSUSY pair of partner Hamiltonians
\begin{equation}
  \hat{H}'_{1,2} = \hat{\pi}^2 + V'_{1,2}(x) + E'_0, \qquad V'_{1,2}(x) = W^{\prime 2}(x) \mp f(x) 
  \frac{dW'}{dx},  \label{eq:V'1,2}
\end{equation}
where
\begin{equation}
  V'_1(x) + E'_0 = V_2(x) + E_0.  \label{eq:V'-V}
\end{equation}
From the ground state wavefunction of $\hat{H}'_1 = \hat{H}_2$, given by
\begin{equation}
  \psi'_0(x) \propto f^{-1/2} \exp\left(- \int^x \frac{W'(x')}{f(x')} dx'\right),  \label{eq:psi'0}
\end{equation}
and corresponding to energy $E'_0$, the first excited state wavefunction of $\hat{H}_1$ with energy $E_1 = E'_0$ is then obtained through the equation
\begin{equation}
  \psi_1(x) \propto \hat{A}^+ \psi'_0(x).  \label{eq:psi1}
\end{equation}
\par
%
%
In terms of the superpotentials, Eq.~(\ref{eq:V'-V}) can be rewritten as
\begin{equation}
  W^2(x) + f(x) \frac{dW}{dx} = W^{\prime 2}(x) - f(x) \frac{dW'}{dx} + E_1 - E_0.  \label{eq:W-W'}
\end{equation}
Such a condition can be satisfied, in particular, whenever, up to some additive constant $R$, $V_1(x)$ and $V_2(x)$ are similar in shape and differ only in the parameters that appear in them. It can then be generalized to any neighbouring members of a DSUSY hierarchy and is known as the DSI condition. In such a special case, the whole bound state spectrum of Eq.~(\ref{eq:def-SE}) (or (\ref{eq:PDM-SE})) can be exactly derived. Several pairs of deforming function $f(x)$ and potential $V(x)$ for which such a situation occurs have been listed in Refs.~\cite{bagchi, cq09}, together with their bound state energies and wavefunctions. In Section 3, we plan to build QES extensions with two known eigenstates of these ES potentials by starting from Eq.~(\ref{eq:W-W'}).\par
%
%
\section{Generating function method for PDM Schr\"odinger equations with two known eigenstates}

\setcounter{equation}{0}

{}From the two superpotentials $W(x)$ and $W'(x)$, assumed to satisfy Eq.~(\ref{eq:W-W'}), let us define the two functions
\begin{equation}
  W_+(x) = W'(x) + W(x), \qquad W_-(x) = W'(x) - W(x).  \label{eq:W_+-W_-}
\end{equation}
In terms of them, Eq.~(\ref{eq:W-W'}) can be rewritten as
\begin{equation}
  f(x) \frac{dW_+}{dx} = W_+(x) W_-(x) + E_1 - E_0,
\end{equation}
thus showing that $W_-(x)$ can be expressed in terms of $W_+(x)$ and the energy difference $E_1-E_0$ as
\begin{equation}
  W_-(x) = \frac{f(x) dW_+(x)/dx + E_0 - E_1}{W_+(x)}.  \label{eq:W_-}
\end{equation}
\par
%
%
The generating function method starts from two functions $W_+(x)$ and $W_-(x)$ that are compatible, i.e., such that Eq.~(\ref{eq:W_-}) is satisfied for some (yet unknown) positive constant $E_1-E_0$. Then the two superpotentials $W(x)$ and $W'(x)$ are determined from Eq.~(\ref{eq:W_+-W_-}). The starting potential $V_1(x)$ [as well as its partner $V_2(x)$] is obtained from Eq.~(\ref{eq:V1,2}) and its ground state wavefunction is derived from Eq.~(\ref{eq:psi0}). For its first excited state wavefunction, Eqs.~(\ref{eq:A}), (\ref{eq:psi'0}), (\ref{eq:psi1}), and (\ref{eq:W_+-W_-}) are combined to yield
\begin{align}
  \psi_1(x) &\propto \left(- f \frac{d}{dx} - \frac{1}{2} \frac{df}{dx} + W\right) f^{-1/2} \exp\left(-
     \int^x \frac{W'(x')}{f(x')} dx'\right) \nonumber \\
  &\propto [W'(x) + W(x)] f^{-1/2} \exp\left(- \int^x \frac{W'(x')}{f(x')} dx'\right) \nonumber \\
  &\propto W_+(x) f^{-1/2} \exp\left(- \int^x \frac{W'(x')}{f(x')} dx'\right). \label{eq:psi1-W+}
\end{align}
The construction of the first two bound state wavefunctions is of course valid provided these functions are square integrable on $(x_1,x_2)$ and fulfil the additional condition (\ref{eq:Hermiticity}).\par
%
%
The difficulty of this generating function method is to guess an appropriate function $W_+(x)$ [and its accompanying function $W_-(x)$], such that Eq.~(\ref{eq:W_-}) is satisfied for some positive constant $E_1-E_0$ and that the potential and wavefunctions derived from it are well behaved. To solve this problem, we are going to combine the method with another one, as explained on an example in Section~4.\par 
%
%
\section{Conditionally deformed shape invariance method: the case of linear harmonic oscillator extensions}

\setcounter{equation}{0}

In Refs.~\cite{bagchi, cq09}, it was shown that the linear harmonic oscillator (HO) potential $V(x) = \frac{1}{4} \omega^2 x^2$, with $\omega>0$ and $-\infty<x<\infty$, remains ES in a PDM background characterized by a deforming function $f(x) = 1 + \alpha x^2$ with $\alpha>0$. To the rescaled potential $V_1(x)$ corresponds the superpotential $W(x) = \lambda x$, where $\lambda = \frac{1}{2}(\alpha + \Delta)$ and $\Delta = \sqrt{\omega^2 + \alpha^2}$, and to its partner $V_2(x) = [\frac{1}{4} \omega^2 + \alpha(\alpha + \Delta)] x^2 - E_0 + \alpha + \Delta$, one can associate the superpotential $W'(x) = (\lambda + \alpha) x$. The bound state energies and wavefunctions are given by
\begin{equation}
  E_n = \left(n + \frac{1}{2}\right) \Delta + \left(n^2 + n + \frac{1}{2}\right) \alpha,
\end{equation}
and
\begin{equation}
  \psi_n(x) \propto f^{-\frac{1}{4}\left(3 + \frac{\Delta}{\alpha}\right)} C_n^{\left(\frac{1}{2} + 
  \frac{\Delta}{2\alpha}\right)}\left(\sqrt{\frac{\alpha}{f}} x\right),
\end{equation}
where $n=0$, 1, 2,~\ldots, and $C_n^{(a)}(t)$ denotes a Gegenbauer polynomial.\par
%
%
Corresponding to the same deforming function $f(x) = 1 + \alpha x^2$ (with $\alpha>0$), let us now consider an infinite set of extensions of the linear HO
\begin{equation}
  V^{(m)}(x) = \sum_{k=1}^{2m+1} B_{2k} x^{2k}, \qquad m=1, 2, \ldots,  \label{eq:Vm}
\end{equation}
with $B_{4m+2} > 0$. As it is obvious, the $m=0$ case would give back $V(x)$ with $B_2 = \frac{1}{4} \omega^2$. For any $m \in \N^+$, we would like to determine the potential parameters $B_2$, $B_4$,~\ldots,
$B_{4m+2}$ in order that the PDM Schr\"odinger equation has known ground and first excited states. We will consider the $m=1$ and $m=2$ cases in detail, then generalize the results to any $m$ by combining this approach with the generating function method of Section~3.\par
%
%
\subsection{\boldmath Conditionally deformed shape invariance method for $m=1$}

{}For $m=1$, we take
\begin{equation}
  V(x) = V^{(1)}(x) = B_2 x^2 + B_4 x^4 + B_6 x^6, \qquad B_6>0,
\end{equation}
in (\ref{eq:def-SE}) and assume a superpotential of the form
\begin{equation}
  W(x) = \lambda x + \mu x^3, \qquad \mu>0,  \label{eq:W}
\end{equation}
depending on two parameters $\lambda$ and $\mu$. From the Riccati equation $V_1(x) = W^2 - f dW/dx$, valid for the rescaled potential (\ref{eq:V1}), it follows that the two unknowns $\lambda$ and $\mu$ satisfy the system of equations
\begin{equation}
\begin{split}
  & - \lambda = - E_0, \\
  & \lambda(\lambda - \alpha) - 3\mu = B_2, \\
  & \mu(2\lambda - 3\alpha) = B_4, \\
  & \mu^2 = B_6.
\end{split}
\end{equation}
The last two equations lead to the values of the unknowns
\begin{equation}
  \mu = \sqrt{B_6}, \qquad \lambda = \frac{B_4}{2\sqrt{B_6}} + \frac{3}{2} \alpha,  \label{eq:lambda-mu}
\end{equation}
while the first equation yields the ground state energy
\begin{equation}
  E_0 = \lambda = \frac{B_4}{2\sqrt{B_6}} + \frac{3}{2} \alpha.
\end{equation}
There remains the second equation, which provides a constraint relating the potential parameters
\begin{equation}
  B_2 = \frac{B_4^2}{4B_6} + \frac{B_4}{\sqrt{B_6}} \alpha + \frac{3}{4} \alpha^2 - 3\sqrt{B_6}.
  \label{eq:constraint}
\end{equation}
To $E_0$ corresponds the ground state wavefunction (\ref{eq:psi0}), which can be rewritten as
\begin{equation}
  \psi_0(x) \propto f^{-\frac{1}{2}\left(1 + \frac{\lambda}{\alpha} - \frac{\mu}{\alpha^2}\right)}
  \exp\left(- \frac{\mu}{2\alpha} x^2\right), \label{eq:psi0-bis}
\end{equation}
where $\lambda$ and $\mu$ are given in Eq.~(\ref{eq:lambda-mu}).\par
%
%
The partner $V_2(x)$ of $V_1(x)$ in DSUSY is, according to (\ref{eq:V1,2}) and (\ref{eq:W}),
\begin{equation}
  V_2(x) = B'_2 x^2 + B'_4 x^4 + B'_6 x^6 + R,
\end{equation}
where $B'_2$, $B'_4$, $B'_6$ are some new parameters and $R$ is a constant. The latter satisfy the equations
\begin{equation}
\begin{split}
  & R = \lambda, \\
  & B'_2 = \lambda(\lambda+\alpha) + 3 \mu, \\
  & B'_4 = \mu(2\lambda+3\alpha), \\
  & B'_6 = \mu^2.
\end{split}
\end{equation}
On using some previous results, we get
\begin{equation}
  B'_2 = B_2 + \alpha\left(\frac{B_4}{\sqrt{B_6}} + 3\alpha\right) + 6 \sqrt{B_6}, \qquad B'_4 = B_4 +
  6\alpha \sqrt{B_6}, \qquad B'_6 = B_6,  \label{eq:B'}
\end{equation}
as well as
\begin{equation}
  R = -E_0 + \frac{B_4}{\sqrt{B_6}} + 3\alpha.
\end{equation}
Hence, the starting potential $V_1(x)$ is DSI, but this deformed shape invariance is not unconditionally valid since constraint (\ref{eq:constraint}) must be satisfied. We may therefore say that the potential is CDSI.\par
%
%
Let us now try to repeat the procedure by taking $V_2(x)$ as a starting potential $V'(x) = V_2(x)$ with ground state energy $E'_0$ and let us represent the new rescaled potential $V'_1(x) = V'(x) + E_0 - E'_0$ (see Eq.~(\ref{eq:V'-V})) as in (\ref{eq:V'1,2}) with
\begin{equation}
  W'(x) = \lambda' x + \mu' x^3, \qquad \mu'>0.  \label{eq:W'}
\end{equation}
By proceeding as in the first step, we obtain for the new parameters
\begin{equation}
  \mu' = \sqrt{B_6}, \qquad \lambda' = \lambda + 3\alpha = \frac{B_4}{2\sqrt{B_6}} + \frac{9}{2}\alpha,
  \label{eq:lambda'}
\end{equation}
and for the partner ground state energy
\begin{equation}
  E'_0 = 3(\lambda + \alpha) = \frac{3B_4}{2\sqrt{B_6}} + \frac{15}{2}\alpha.
\end{equation}
There also occurs a new constraint coming from the relation $B'_2 = \lambda'(\lambda'-\alpha) - 3\mu'$. On taking Eqs.~(\ref{eq:B'}) and (\ref{eq:lambda'}) into account, it can be written as
\begin{equation}
  B_2 = \frac{B_4^2}{4B_6} + 3\alpha \frac{B_4}{\sqrt{B_6}} + \frac{51}{4}\alpha^2 - 9 \sqrt{B_6}.
  \label{eq:constraint-bis}
\end{equation}
The partner ground state wavefunction $\psi'_0(x)$ is given by an equation similar to (\ref{eq:psi0-bis}) with $\lambda'$, $\mu'$ substituted for $\lambda$, $\mu$.\par
%
%
The two constraints (\ref{eq:constraint}) and (\ref{eq:constraint-bis}) are compatible provided
\begin{equation}
  B_4 = \frac{3B_6}{\alpha} - 6\alpha \sqrt{B_6}. 
\end{equation}
Then
\begin{equation}
  B_2 = \frac{9B_6}{4\alpha^2} - 9 \sqrt{B_6} + \frac{15}{4}\alpha^2
\end{equation}
and
\begin{equation}
  \lambda = \frac{3\sqrt{B_6}}{2\alpha} - \frac{3}{2}\alpha, \qquad \lambda' = \frac{3\sqrt{B_6}}{2\alpha} +
  \frac{3}{2}\alpha. 
\end{equation}
Hence the potential
\begin{equation}
  V(x) = \left(\frac{9B_6}{4\alpha^2} - 9\sqrt{B_6} + \frac{15}{4}\alpha^2\right) x^2 + \left(\frac{3B_6}
  {\alpha} - 6\alpha\sqrt{B_6}\right) x^4 + B_6 x^6, \qquad B_6>0,
\end{equation}
with corresponding superpotentials
\begin{equation}
  W(x) = \left(\frac{3\sqrt{B_6}}{2\alpha} - \frac{3}{2}\alpha\right) x + \sqrt{B_6} x^3, \qquad
  W'(x) = \left(\frac{3\sqrt{B_6}}{2\alpha} + \frac{3}{2}\alpha\right) x + \sqrt{B_6} x^3,
  \label{eq:W-m=1}
\end{equation}
in a PDM background characterized by the mass $m(x) = 1/(1+\alpha x^2)^2$ (with $\alpha>0$), has a ground state and a first excited state, whose energies are given by
\begin{equation}
  E_0 = \frac{3\sqrt{B_6}}{2\alpha} - \frac{3}{2}\alpha, \qquad E_1 = E'_0 = \frac{9\sqrt{B_6}}{2\alpha}
  - \frac{3}{2}\alpha,  \label{eq:E-E}
\end{equation}
respectively.\par
%
%
The ground state wavefunction of this potential is given by
\begin{equation}
  \psi_0(x) \propto f^{-\frac{1}{4}\left(\frac{\sqrt{B_6}}{\alpha^2} - 1\right)} \exp\left(- \frac{\sqrt{B_6}}
  {2\alpha} x^2\right),  \label{eq:psi0-m=1}
\end{equation}
while the first excited state wavefunction can be obtained from the partner ground state wavefunction
\begin{equation}
  \psi'_0(x) \propto f^{-\frac{1}{4}\left(\frac{\sqrt{B_6}}{\alpha^2} + 5\right)} \exp\left(- \frac{\sqrt{B_6}}
  {2\alpha} x^2\right),
\end{equation}
by acting with the operator $\hat{A}^+$, given in (\ref{eq:A}) and (\ref{eq:W-m=1}). The result reads
\begin{equation}
  \psi_1(x) \propto f^{-\frac{1}{4}\left(\frac{\sqrt{B_6}}{\alpha^2}+5\right)} (3x + 2\alpha x^3)
  \exp\left(- \frac{\sqrt{B_6}}{2\alpha} x^2\right),  \label{eq:psi1-m=1}
\end{equation}
which has a single zero at $x=0$, as it should be. Both wavefunctions (\ref{eq:psi0-m=1}) and (\ref{eq:psi1-m=1}) are square integrable on the real line and satisfy the additional condition (\ref{eq:Hermiticity}) for $x \to \pm \infty$.\par
%
%
{}From the two superpotentials (\ref{eq:W-m=1}), we obtain for the functions $W_{\pm}(x)$ of Section~3 the following results:
\begin{equation}
  W_+(x) = \frac{3\sqrt{B_6}}{\alpha} x + 2 \sqrt{B_6} x^3, \qquad W_-(x) = 3\alpha x.
\end{equation}
It can be easily checked that these functions satisfy Eq.~(\ref{eq:W_-}) with $E_1-E_0 = \frac{3\sqrt{B_6}}{\alpha}$, in accordance with Eq.~(\ref{eq:E-E}). Furthermore, Eq.~(\ref{eq:psi1-m=1}) also agrees with Eq.~(\ref{eq:psi1-W+}), as it should be.\par
%
%
\subsection{\boldmath Conditionally deformed shape invariance method for $m=2$}

Turning now ourselves  to the $m=2$ case of Eq.~(\ref{eq:Vm}), corresponding to the potential
\begin{equation}
  V(x) = V^{(2)}(x) = B_2 x^2 + B_4 x^4 + B_6 x^6 + B_8 x^8 + B_{10} x^{10}, \qquad B_{10}>0,
  \label{eq:V-m=2}
\end{equation}
considering a superpotential of the type
\begin{equation}
  W(x) = \lambda x + \mu x^3 + \nu x^5, \qquad \nu>0,
\end{equation}
and proceeding as in Section~4.1, we obtain
\begin{equation}
  \nu = \sqrt{B_{10}}, \qquad \mu = \frac{B_8}{2\sqrt{B_{10}}}, \qquad \lambda = E_0 = 
  \frac{1}{2\sqrt{B_{10}}} \left(B_6 - \frac{B_8^2}{4B_{10}}\right) + \frac{5}{2}\alpha,
  \label{eq:lambda-mu-nu}
\end{equation}
as well as two constraints
\begin{equation}
  B_2 = \frac{1}{4B_{10}}\left(B_6 - \frac{B_8^2}{4B_{10}}\right)^2 + \frac{2\alpha}{\sqrt{B_{10}}}
  \left(B_6 - \frac{B_8^2}{4B_{10}}\right) + \frac{15}{4} \alpha^2 - \frac{3B_8}{2\sqrt{B_{10}}},
  \label{eq:constraint1}
\end{equation}
\begin{equation}
  B_4 = \frac{B_8}{2\sqrt{B_{10}}} \left[\frac{1}{\sqrt{B_{10}}}\left(B_6 - \frac{B_8^2}{4B_{10}}\right)
  + 2\alpha\right] - 5 \sqrt{B_{10}},  \label{eq:constraint2}
\end{equation}
because there are now two more parameters in the potential than in the superpotential. The ground state wavefunction reads
\begin{equation}
  \psi_0(x) \propto f^{-\frac{1}{2}\left(1 + \frac{\lambda}{\alpha} - \frac{\mu}{\alpha^2} + 
  \frac{\nu}{\alpha^3}\right)} \exp\left((\nu-\alpha\mu) \frac{x^2}{2\alpha^2} - \nu\frac{x^4}{4\alpha}\right),
  \label{eq:psi0-ter}
\end{equation}
with $\lambda$, $\mu$, and $\nu$ given in (\ref{eq:lambda-mu-nu}).\par
%
%
The partner $V_2(x)$ of $V_1(x) = V(x) - E_0$ is given by
\begin{equation}
  V_2(x) = B'_2 x^2 + B'_4 x^4 + B'_6 x^6  + B'_8 x^8 + B'_{10} x^{10} + R, 
\end{equation}
where
\begin{equation}
\begin{split}
  &B'_2 = B_2 + \frac{\alpha}{\sqrt{B_{10}}}\left(B_6 - \frac{B_8^2}{4B_{10}}\right) + 5\alpha^2
      + \frac{3B_8}{\sqrt{B_{10}}}, \\ 
  &B'_4 = B_4 + 3\alpha \frac{B_8}{\sqrt{B_{10}}} + 10 \sqrt{B_{10}},\qquad B'_6 = B_6 + 
      10\alpha\sqrt{B_{10}}, \\ 
  &B'_8 = B_8, \qquad B'_{10} = B_{10}, \qquad R = - E_0 + \frac{1}{\sqrt{B_{10}}}\left(B_6 - 
      \frac{B_8^2}{4B_{10}}\right) + 5\alpha,
\end{split}
\end{equation}
showing that $V_1(x)$ is CDSI with constraints (\ref{eq:constraint1}) and (\ref{eq:constraint2}).\par
%
%
The superpotential $W'(x)$ of Eq.~(\ref{eq:W'}) is now replaced by
\begin{equation}
  W'(x) = \lambda' x + \mu' x^3 + \nu' x^5, \qquad \nu'>0.
\end{equation}
The latter leads to the relations
\begin{equation}
\begin{split}
  &\nu' = \nu = \sqrt{B_{10}}, \qquad \mu' = \mu = \frac{B_8}{2\sqrt{B_{10}}}, \\ 
  &\lambda' = \lambda + 5\alpha = \frac{1}{2\sqrt{B_{10}}}\left(B_6 - \frac{B_8^2}{4B_{10}}\right) +
      \frac{15}{2}\alpha, \\
  &E'_0 = 3\lambda + 5\alpha = \frac{3}{2\sqrt{B_{10}}}\left(B_6 - \frac{B_8^2}{4B_{10}}\right) + 
     \frac{25}{2}\alpha,
\end{split}
\end{equation}
and to the two new constraints
\begin{equation}
  B_2 = \frac{1}{4B_{10}}\left(B_6 - \frac{B_8^2}{4B_{10}}\right)^2 + \frac{6\alpha}{\sqrt{B_{10}}}
  \left(B_6 - \frac{B_8^2}{4B_{10}}\right) + \frac{175}{4} \alpha^2 - \frac{9B_8}{2\sqrt{B_{10}}},
  \label{eq:constraint1-bis}
\end{equation}
\begin{equation}
  B_4 = \frac{B_8}{2\sqrt{B_{10}}} \left[\frac{1}{\sqrt{B_{10}}}\left(B_6 - \frac{B_8^2}{4B_{10}}\right)
  + 6\alpha\right] - 15 \sqrt{B_{10}}.  \label{eq:constraint2-bis}
\end{equation}
The partner ground state wavefunction $\psi'_0(x)$ is similar to (\ref{eq:psi0-ter}) with all parameters replaced by primed ones.\par
%
%
The two sets of constraints (\ref{eq:constraint1}), (\ref{eq:constraint2}), and (\ref{eq:constraint1-bis}), (\ref{eq:constraint2-bis}) are compatible provided
\begin{equation}
  B_8 = \frac{5B_{10}}{\alpha}, \qquad B_6 = 10\sqrt{B_{10}}\left(\frac{\sqrt{B_{10}}}{\alpha^2} -
  \alpha\right).  \label{eq:parameters1}
\end{equation}
We then also obtain
\begin{equation}
  B_4 = \frac{25}{2}\sqrt{B_{10}}\left(\frac{3\sqrt{B_{10}}}{4\alpha^3} - 2\right), \qquad
  B_2 = \frac{5}{4}\left(\frac{45B_{10}}{16\alpha^4} - \frac{15\sqrt{B_{10}}}{\alpha} + 7\alpha^2\right),
  \label{eq:parameters2}
\end{equation}
and
\begin{equation}
\begin{split}
  &\nu = \nu' = \sqrt{B_{10}}, \qquad \mu = \mu' = \frac{5\sqrt{B_{10}}}{2\alpha}, \\
  & \lambda = \frac{5}{2}\left(\frac{3\sqrt{B_{10}}}{4\alpha^2} - \alpha\right), \qquad \lambda' =
  \frac{5}{2}\left(\frac{3\sqrt{B_{10}}}{4\alpha^2} + \alpha\right).
\end{split}
\end{equation}
We conclude that the potential (\ref{eq:V-m=2}), with parameters given in (\ref{eq:parameters1}), (\ref{eq:parameters2}), and corresponding superpotentials
\begin{equation}
\begin{split}
  &W(x) = \frac{5}{2}\left(\frac{3\sqrt{B_{10}}}{4\alpha^2} - \alpha\right) x + \frac{5\sqrt{B_{10}}}{2\alpha}
     x^3 + \sqrt{B_{10}} x^5, \\
  &W'(x) = \frac{5}{2}\left(\frac{3\sqrt{B_{10}}}{4\alpha^2} + \alpha\right) x + \frac{5\sqrt{B_{10}}}
    {2\alpha} x^3 + \sqrt{B_{10}} x^5,
\end{split} \label{eq:W-m=2}
\end{equation}
has a ground state and a first excited state, whose energies are given by
\begin{equation}
  E_0 = \frac{5}{2}\left(\frac{3\sqrt{B_{10}}}{4\alpha^2} - \alpha\right), \qquad E_1 = E'_0 = 
  \frac{5}{2}\left(\frac{9\sqrt{B_{10}}}{4\alpha^2} - \alpha\right), \label{eq:E-E-bis}
\end{equation}
respectively. The corresponding wavefunctions read
\begin{equation}
\begin{split}
  &\psi_0(x) \propto f^{-\frac{3}{16}\left(\frac{\sqrt{B_{10}}}{\alpha^3} - 4\right)} \exp{\left(-
     \frac{3\sqrt{B_{10}}}{4\alpha^2} x^2 - \frac{\sqrt{B_{10}}}{4\alpha} x^4\right)}, \\
  &\psi_1(x) \propto f^{-\frac{1}{16}\left(\frac{3\sqrt{B_{10}}}{\alpha^3} + 28\right)} (15x + 20\alpha x^3
     + 8\alpha^2 x^5) \exp{\left(-\frac{3\sqrt{B_{10}}}{4\alpha^2} x^2 - \frac{\sqrt{B_{10}}}{4\alpha}
     x^4\right)},
\end{split}
\end{equation}
and satisfy both the square integrability condition and that given in Eq.~(\ref{eq:Hermiticity}). It is also obvious that $\psi_1(x)$ has a single zero at $x=0$, as it should be.\par
%
%
{}From the two superpotentials (\ref{eq:W-m=2}), we now obtain
\begin{equation}
  W_+(x) = \frac{15\sqrt{B_{10}}}{4\alpha^2} x + \frac{5\sqrt{B_{10}}}{\alpha} x^3 + 2 \sqrt{B_{10}} x^5,
  \qquad W_-(x) = 5\alpha x,
\end{equation}
fulfilling Eq.~(\ref{eq:W_-}) with $E_1 - E_0 = \frac{15\sqrt{B_{10}}}{4\alpha^2}$, in accordance with Eq.~(\ref{eq:E-E-bis}).\par
%
%
\subsection{\boldmath Results for any $m \in \N^+$}

It is obvious that we might continue applying the CDSI method to potentials $V(x) = V^{(m)}(x)$ with $m$ values higher than two. However, since the number of constraints would grow with $m$, the complexity of the method would increase correspondingly. It is then advantageous to turn ourselves to the generating function method of Section~3 by using the fact that we already know $W_+(x)$ and $W_-(x)$ for $m=1$ and $m=2$. As a generalization of the latter for any $m \in \N^+$, we propose to consider
\begin{equation}
  W_+(x) = 2\sqrt{B_{4m+2}} \sum_{k=0}^m \frac{(2m+1)!!}{(2k+1)!! (2m-2k)!!} \alpha^{k-m} x^{2k+1},
  \qquad W_-(x) = (2m+1)\alpha x.  \label{eq:W+W-m}
\end{equation}
To be acceptable, this choice must be such that Eq.~(\ref{eq:W_-}) is satisfied for some positive constant $E_1 - E_0$. It is straightforward to show that this is indeed so for
\begin{equation}
  E_1 - E_0 = \frac{2 (2m+1)!! \sqrt{B_{4m+2}}}{(2m)!! \alpha^m}. \label{eq:E-E-m}
\end{equation}
This result agrees with those found in Sections 4.1 and 4.2 for $m=1$ and $m=2$, respectively.\par
%
%
{}From Eq.~(\ref{eq:W+W-m}), we obtain the two superpotentials
\begin{align}
  W(x) &= \left[\frac{(2m+1)!!}{(2m)!!} \alpha^{-m} \sqrt{B_{4m+2}} - \left(m + \frac{1}{2}\right) \alpha
      \right] x \nonumber \\
  &\quad + \sqrt{B_{4m+2}} \sum_{k=1}^m \frac{(2m+1)!!}{(2k+1)!! (2m-2k)!!} \alpha^{k-m} x^{2k+1},
      \label{eq:Wm} 
\end{align}
\begin{align}
  W'(x) &= \left[\frac{(2m+1)!!}{(2m)!!} \alpha^{-m} \sqrt{B_{4m+2}} + \left(m + \frac{1}{2}\right) \alpha
      \right] x \nonumber \\
  &\quad + \sqrt{B_{4m+2}} \sum_{k=1}^m \frac{(2m+1)!!}{(2k+1)!! (2m-2k)!!} \alpha^{k-m} x^{2k+1}. 
\end{align}
\par
%
%
Equations (\ref{eq:V1}), (\ref{eq:V1,2}), and (\ref{eq:Wm}) now allow us to determine the expansion coefficients $B_{2k}$, $k=1$, 2,~\ldots, $2m+1$, in Eq.~(\ref{eq:Vm}), as well as the ground state energy $E_0$. They can be written as
\begin{equation}
\begin{split}
  B_2 &= \left(m+\frac{1}{2}\right)\left(m+\frac{3}{2}\right) \alpha^2 - 2 (2m+1) \frac{(2m+1)!!}{(2m)!!} 
      \alpha^{-m+1} \sqrt{B_{4m+2}} \\
  &\quad + \left(\frac{(2m+1)!!}{(2m)!!}\right)^2 \alpha^{-2m} B_{4m+2}, \\
  B_{2k} &= - \frac{2(2m+1) (2m+1)!!}{(2k-1)!! (2m-2k+2)!!} \alpha^{k-m} \sqrt{B_{4m+2}} 
      + S^{(m,k)}_{0,k-1} \alpha^{k-2m-1} B_{4m+2}, \\
  &\qquad k=2, 3, \ldots, m+1, \\
  B_{2k} &= S^{(m,k)}_{k-m-1,m} \alpha^{k-2m-1} B_{4m+2}, \qquad k=m+2, m+3, \ldots, 2m,
\end{split}
\end{equation}
and
\begin{equation}
  E_0 = - \left(m+\frac{1}{2}\right) \alpha + \frac{(2m+1)!!}{(2m)!!} \alpha^{-m} \sqrt{B_{4m+2}},
  \label{eq:E0-m}
\end{equation}
where we have defined
\begin{equation}
  S^{(m,k)}_{a,b} = \sum_{l=a}^b \frac{[(2m+1)!!]^2}{(2l+1)!! (2k-2l-1)!! (2m-2l)!! (2m-2k+2l+2)!!}.
\end{equation}
The first excited state energy $E_1$ directly follows from Eqs.~(\ref{eq:E-E-m}), (\ref{eq:E0-m}), and is given by
\begin{equation}
  E_1 = - \left(m+\frac{1}{2}\right) \alpha + 3 \frac{(2m+1)!!}{(2m)!!} \alpha^{-m} \sqrt{B_{4m+2}}.
\end{equation}
\par
%
%
In the general $m$ case, the partner $V_2(x) = V_1(x) + 2f(x) dW/dx$ of $V_1(x)$ can be written as
\begin{equation}
  V_2(x) = \sum_{k=1}^{2m+1} B'_{2k} x^{2k} + R,
\end{equation}
where
\begin{equation}
\begin{split}
  B'_2 &= \left(m-\frac{1}{2}\right) \left(m+\frac{1}{2}\right) \alpha^2 + \left(\frac{(2m+1)!!}{(2m)!!}
      \right)^2 \alpha^{-2m} B_{4m+2}, \\
  B'_{2k} &= S^{(m,k)}_{0,k-1} \alpha^{k-2m-1} B_{4m+2}, \qquad k=2, 3, \ldots, m+1, \\
  B'_{2k} &= S^{(m,k)}_{k-m-1,m} \alpha^{k-2m-1} B_{4m+2}, \qquad k=m+2, m+3, \ldots, 2m+1, \\
  R &= - \left(m+\frac{1}{2}\right) \alpha + \frac{(2m+1)!!}{(2m)!!} \alpha^{-m} \sqrt{B_{4m+2}}. 
\end{split}
\end{equation}
\par
%
%
The wavefunctions of $V(x)$, corresponding to energies $E_0$ and $E_1$, are given by
\begin{align}
  \psi_0(x) &\propto f^{-\frac{1}{2} \left(\frac{1}{2} - m + \frac{(2m-1)!!}{(2m)!!} \alpha^{-m-1}
      \sqrt{B_{4m+2}}\right)} \nonumber \\
  &\quad \times \exp{\left(- \frac{1}{2} \sqrt{B_{4m+2}} \sum_{l=1}^m \frac{(2m-1)!!}{(2l-1)!! (2m-2l)!!}
      \alpha^{l-m-1} \frac{x^{2l}}{l}\right)},  \label{eq:psi0-m}
\end{align}
and
\begin{align}
  \psi_1(x) &\propto f^{-\frac{1}{2} \left(\frac{3}{2} + m + \frac{(2m-1)!!}{(2m)!!} \alpha^{-m-1}
      \sqrt{B_{4m+2}}\right)} \left(\sum_{k=0}^m \frac{(2m+1)!!}{(2k+1)!! (2m-2k)!!} \alpha^k x^{2k+1}
      \right) \nonumber \\
  &\quad \times \exp{\left(- \frac{1}{2} \sqrt{B_{4m+2}} \sum_{l=1}^m \frac{(2m-1)!!}{(2l-1)!! (2m-2l)!!}
      \alpha^{l-m-1} \frac{x^{2l}}{l}\right)},  \label{eq:psi1-m}
\end{align}
respectively. As it can be easily checked, both wavefunctions (\ref{eq:psi0-m}) and (\ref{eq:psi1-m}) are physically acceptable.\par
%
%
In Fig.~1, an example of extended HO potential is plotted. Its corresponding (unnormalized) wavefunctions $\psi_0(x)$ and $\psi_1(x)$ are displayed in Fig.~2.\par
%
%
\section{Other examples of extension families}

We have applied the method presented in Section~4 to other pairs of DSI potential and deforming function, derived in Refs.~\cite{bagchi, cq09}. They correspond to the $d$-dimensional radial harmonic oscillator (RHO), the $d$-dimensional radial Kepler-Coulomb (KC), and the Morse potential (M). They are listed in Table~1, together with their associated superpotential, while their bound state energies are given in Table~2. It may be observed that, as in the constant mass case, the radial harmonic oscillator and the Morse potential have an infinite or finite number of bound states, respectively. In contrast, the infinite number of bound states of the constant mass Kepler-Coulomb potential is converted into a finite one, showing the strong influence the presence of a PDM may have on the spectrum.\par
%
%
\begin{table}[h!]

\caption{Starting potentials, deforming functions, and superpotentials}

\begin{center}
\begin{tabular}{llll}
  \hline\hline\\[-0.2cm]
  Type & $V(x)$ & $f(x)$ & $W(x)$ \\[0.2cm]
  \hline\\[-0.2cm]
  RHO & $\frac{L(L+1)}{x^2} + \frac{1}{4} \omega^2 x^2$ & $1 + \alpha x^2$ & $\frac{\lambda}{x} + \mu x$
      \\[0.2cm]
  & $0<x<\infty$, $L = l + \frac{d-3}{2}$, $\omega>0$ & $\alpha>0$ & $\lambda = -L-1$, $\mu = 
      \frac{1}{2}(\alpha+\Delta)$ \\[0.2cm] 
  & & & $\Delta = \sqrt{\omega^2 + \alpha^2}$ \\[0.2cm]
  KC & $\frac{L(L+1)}{x^2} - \frac{Q}{x}$ & $1 + \alpha x$ & $\frac{\lambda}{x} + \mu$ \\[0.2cm]
  & $0<x<\infty$, $L = l + \frac{d-3}{2}$, $Q>0$ & $\alpha>0$ & $\lambda = -L-1$, $\mu = \frac{Q - \alpha
      (L+1)}{2(L+1)}$ \\[0.2cm]
  M & $B^2 e^{-2x} - B(2A+1) e^{-x}$ & $1 + \alpha e^{-x}$ & $\lambda e^{-x} + \mu$ \\[0.2cm]
  & $-\infty < x < \infty$, $A, B >0$ & $\alpha>0$ & $\lambda = - \frac{1}{2}(\alpha + \Delta)$, $\Delta =
       \sqrt{4B^2 + \alpha^2}$ \\[0.2cm]
  & & & $\mu = - \frac{1}{2}\left(\frac{B(2A+1)}{\lambda} + 1\right)$ \\[0.2cm]
  \hline \hline
\end{tabular}
\end{center}

\end{table}
\par
%
%
\begin{table}[h!]

\caption{Bound state energy spectra of starting potentials}

\begin{center}
\begin{tabular}{lll}
  \hline\hline\\[-0.2cm]
  Type & $E_n$ & $n$ \\[0.2cm]
  \hline\\[-0.2cm]
  RHO & $\Delta \left(2n + L + \frac{3}{2}\right)$ & $0, 1, 2, \ldots$ \\[0.2cm]
  & $+ \alpha \left[2(n+L+1)(2n+1) + \frac{1}{2}\right]$ & \\[0.2cm]
  KC & $- \left(\frac{Q - \alpha[n^2 + (L+1)(2n+1)]}{2(n+L+1)}\right)^2$ & $0, 1, \ldots, n_{\rm max}$
      \\[0.2cm]
  & & $n_{\rm max}^2 + (L+1) (2n_{\rm max}+1) < \frac{Q}{\alpha}$ \\[0.2cm]
  & & $\le (n_{\max}+1)^2 + (L+1) (2n_{\rm max}+3)$ \\[0.2cm]
  M & $- \frac{1}{4} \left(\frac{2B(2A+1) - [(2n+1)\Delta + (2n^2+2n+1)\alpha]}{\Delta + (2n+1)\alpha}
       \right)^2$ & $0, 1, \ldots, n_{\rm max}$ \\[0.2cm] 
  & & $n_{\rm max}(2|\lambda| + n_{\rm max}\alpha) < 2 |\lambda| \mu$ \\[0.2cm]
  & & $ \le (n_{\rm max}+1) [2|\lambda|+ (n_{\rm max}+1) \alpha]$ \\[0.2cm]
  \hline \hline
\end{tabular}
\end{center}

\end{table}
\par
%
%
The families of extensions we have considered are presented in Table~3, together with the corresponding deforming function. It is obvious that for $m=0$, we would get back the starting potentials of Table~1 with $B_2 = \frac{1}{4}\omega^2$, $B_{-1} = - Q$, and $B_{-2} = B^2$, $B_{-1} = - B(2A+1)$ in the RHO, KC, and M cases, respectively.\par
%
%
\begin{table}[h!]

\caption{Families of extensions and corresponding deforming functions}

\begin{center}
\begin{tabular}{lll}
  \hline\hline\\[-0.2cm]
  Type & $V^{(m)}(x)$ & $f(x)$ \\[0.2cm]
  \hline\\[-0.2cm]
  RHO & $\frac{L(L+1)}{x^2} + \sum_{k=1}^{2m+1} B_{2k} x^{2k}$ & $1 + \alpha x^2$ \\[0.2cm]
  & $0<x<\infty$, $L = l +\frac{d-3}{2}$, $B_{4m+2}>0$ & $\alpha>0$ \\[0.2cm]
  KC & $\frac{L(L+1)}{x^2} + \frac{B_{-1}}{x} + \sum_{k=1}^{2m} B_k x^k$ & $1 + \alpha x$
      \\[0.2cm]
  & $0<x<\infty$, $L = l + \frac{d-3}{2}$, $B_{2m}>0$ & $\alpha>0$ \\[0.2cm]
  M & $B_{-2} e^{-2x} + B_{-1} e^{-x} + \sum_{k=1}^{2m} B_k e^{kx}$ & $1 + \alpha e^{-x}$ \\[0.2cm] 
  & $-\infty<x<\infty$, $B_{-2}, B_{2m}>0$ & $\alpha>0$ \\[0.2cm]
  \hline \hline
\end{tabular}
\end{center}

\end{table}
\par
%
%
The generating functions $W_{\pm}(x)$, corresponding to the extensions of Table~3, are listed in Table~4 and the resulting values of the potential parameters are given in Table~5. It is worth observing that $B_{-1}>0$ for the KC potential extensions whereas, for the starting potential, we have $B_{-1} = -Q<0$. In other words, the extensions obtained are actually those of the repulsive KC potential, which itself has no bound state. A similar change of sign is obtained for the parameter $B_{-1}$ of the Morse potential extensions, which becomes positive instead of $B_{-1} = -B(2A+1) < 0$ for the corresponding starting potential.\par
%
%
\begin{table}[h!]

\caption{Generating functions $W_{\pm}(x)$}

\begin{center}
\begin{tabular}{lll}
  \hline\hline\\[-0.2cm]
  Type & $W_+(x)$ & $W_-(x)$ \\[0.2cm]
  \hline\\[-0.2cm]
  RHO & $- \frac{2L+3}{x} + 2 \sqrt{B_{4m+2}} \sum_{k=1}^{m+1} \binom{m+1}{k} \alpha^{k-m-1} x^{2k-1}$
       & $- \frac{1}{x} + (2m+1) \alpha x$ \\[0.2cm]
  KC & $- \frac{2L+3}{x} - (m+1) \alpha (2L+3)$ & $- \frac{1}{x} + m\alpha $ \\[0.2cm]
  & $+ 2 \sqrt{B_{2m}} \sum_{k=1}^m \binom{m+1}{k+1} \alpha^{k-m} x^k$ & \\[0.2cm]
  M & $- (2\alpha + \Delta) e^{-x} - (m+1) \left(\frac{\Delta}{\alpha} + 2\right)$ & $- \alpha e^{-x} + m$ 
       \\[0.2cm] 
  & $+ 2 \sqrt{B_{2m}} \sum_{k=1}^m \binom{m+1}{k+1} \alpha^{m-k} e^{kx}$ & \\[0.2cm]
  \hline \hline
\end{tabular}
\end{center}

\end{table}
\par
%
%
\begin{table}[h!]

\caption{Parameters of $V^{(m)}(x)$}

\begin{center}
\begin{tabular}{ll}
  \hline\hline\\[-0.2cm]
  Type & Parameters \\[0.2cm]
  \hline\\[-0.2cm]
  RHO & $B_2 = \left(m+\frac{1}{2}\right)\left(m+\frac{3}{2}\right) \alpha^2 - \frac{1}{2} (m+1) (2mL
       +9m+4) \alpha^{1-m} \sqrt{B_{4m+2}}$ \\[0.2cm]
  & $\quad + (m+1)^2 \alpha^{-2m} B_{4m+2}$ \\[0.2cm]
  & $B_{2k} = - \left[\binom{m+1}{k+1} (2L+2k+3) + \binom{m+1}{k} (2m+2k)\right] \alpha^{k-m}
        \sqrt{B_{4m+2}}$ \\[0.2cm]
  & $\quad + \left[\binom{2m+2}{k+1} - 2\binom{m+1}{k+1}\right] \alpha^{k-1-2m} B_{4m+2}$, $k=2, 3,
        \ldots, m+1$ \\[0.2cm]
  & $B_{2k} = \binom{2m+2}{k+1} \alpha^{k-1-2m} B_{4m+2}$, $k=m+2, m+3, \ldots, 2m$ \\[0.2cm]
  KC & $B_{-1} = 2\alpha (L+1) [(m+1)L+2m+1]$ \\[0.2cm]
  & $B_1 = - \frac{2}{3}m(m+1)(2m+1)(L+2) \alpha^{2-m} \sqrt{B_{2m}}$ \\[0.2cm]
  & $B_k = - \left\{\binom{m+1}{k+2}(2L+k+3) + \binom{m+1}{k+1}[(2m+2)L + 4m+k+3]\right\}   
       \alpha^{k-m+1} \sqrt{B_{2m}}$ \\[0.2cm]
  & $\quad + \left[\binom{2m+2}{k+2} - 2\binom{m+1}{k+2} - 2(m+1) \binom{m+1}{k+1}\right] 
       \alpha^{k-2m} B_{2m}$, $k=2, 3, \ldots, m$ \\[0.2cm]
  & $B_k = \binom{2m+2}{k+2} \alpha^{k-2m} B_{2m}$, $k=m+1, m+2, \ldots, 2m-1$ \\[0.2cm]
  M & $B_{-2} = B^2$ \\[0.2cm]
  & $B_{-1} = \frac{1}{2\alpha} (\alpha+\Delta) [(3m+1)\alpha + (m+1) \Delta]$ \\[0.2cm]
  & $B_1 = - \frac{1}{3} m(m+1)(2m+1) (3\alpha+\Delta) \alpha^{m-2} \sqrt{B_{2m}}$ \\[0.2cm]
  & $B_k = - \bigl\{\left[(k+2) \binom{m+1}{k+2} + (3m+k+2) \binom{m+1}{k+1}\right] \alpha$ \\[0.2cm]
  & $\quad + \left[\binom{m+1}{k+2} + (m+1) \binom{m+1}{k+1}\right] \Delta\bigr\} \alpha^{m-k-1}
       \sqrt{B_{2m}}$, \\[0.2cm]
  & $\quad + \left[\binom{2m+2}{k+2} - 2 \binom{m+1}{k+2} - 2(m+1) \binom{m+1}{k+1}\right]
       \alpha^{2m-k} B_{2m}$, $k=2, 3, \ldots, m$ \\[0.2cm]
  & $B_k = \binom{2m+2}{k+2} \alpha^{2m-k} B_{2m}$, $k=m+1, m+2, \ldots, 2m-1$ \\[0.2cm]
  \hline \hline
\end{tabular}
\end{center}

\end{table}
\par
%
%
The ground and first excited state energies of the extensions $V^{(m)}(x)$, given in Tables~3 and 5, are listed in Table~6 and the corresponding wavefunctions  are provided in Tables 7 and 8.\par
%
%
\begin{table}[h!]

\caption{Ground and first excited state energies of $V^{(m)}(x)$}

\begin{center}
\begin{tabular}{lll}
  \hline\hline\\[-0.2cm]
  Type & $E_0$ & $E_1$ \\[0.2cm]
  \hline\\[-0.2cm]
  RHO & $- \left(2mL + 3m + \frac{1}{2}\right) \alpha$ & $\left[(2m+4)L + 3m + \frac{11}{2}\right]
       \alpha$ \\[0.2cm]
  & $+ (m+1) (2L+3) \alpha^{-m} \sqrt{B_{4m+2}}$ & $+ (m+1) (2L+7) \alpha^{-m} \sqrt{B_{4m+2}}$ 
       \\[0.2cm]
  KC & $- \left[(m+1)L + 2m + \frac{3}{2}\right]^2 \alpha^2$ & $- \left[(m+1)L + m + \frac{3}{2}\right]^2
       \alpha^2$ \\[0.2cm]
  & $+ \frac{1}{2} m(m+1)(2L+3) \alpha^{1-m} \sqrt{B_{2m}}$ & $+ \frac{1}{2} m(m+1)(2L+7)
       \alpha^{1-m} \sqrt{B_{2m}}$ \\[0.2cm]
  M & $- \frac{1}{4}\left[(m+1) \frac{\Delta}{\alpha} + 3m+2\right]^2$ & $- \frac{1}{4} \left[(m+1) 
       \frac{\Delta}{\alpha} + m +2\right]^2$ \\[0.2cm]
  & $+ \frac{1}{2} m(m+1)(2\alpha + \Delta) \alpha^{m-1} \sqrt{B_{2m}}$ & $+ \frac{1}{2} m(m+1)
       (6\alpha+\Delta) \alpha^{m-1} \sqrt{B_{2m}}$ \\[0.2cm]
  \hline \hline
\end{tabular}
\end{center}

\end{table}
\par
%
%
\begin{table}[h!]

\caption{Ground state wavefunction of $V^{(m)}(x)$}

\begin{center}
\begin{tabular}{ll}
  \hline\hline\\[-0.2cm]
  Type & $\psi_0(x)$ \\[0.2cm]
  \hline\\[-0.2cm]
  RHO & $x^{L+1} f^{-\frac{1}{2}\left(L-m+\frac{3}{2} + \alpha^{-m-1} \sqrt{B_{4m+2}}\right)}$ \\[0.2cm]
  & $\times \exp\left[- \frac{1}{2} \sqrt{B_{4m+2}} \sum_{l=1}^m \frac{1}{l} \binom{m}{l} \alpha^{l-m-1}
       x^{2l}\right]$ \\[0.2cm]
  KC & $x^{L+1} f^{m\left(L+2 + \alpha^{-m-1} \sqrt{B_{2m}}\right)}$ \\[0.2cm]
  & $\times \exp\left[- \sqrt{B_{2m}} \sum_{l=1}^m \frac{1}{l} \binom{m}{l} \alpha^{l-m-1} x^l \right]$ 
       \\[0.2cm]
  M & $f^{m\left(\frac{3}{2} + \frac{\Delta}{2\alpha} + \alpha^m \sqrt{B_{2m}}\right)} 
       \exp\left\{\left[\frac{1}{2}
       \left((m+1)\frac{\Delta}{\alpha} + 3m+2\right) + m \alpha^m \sqrt{B_{2m}}\right] x\right\}$ \\[0.2cm]
  & $\times \exp\left[- \sqrt{B_{2m}} \sum_{l=1}^m \frac{1}{l} \binom{m}{l} \alpha^{m-l} e^{lx}\right]$ 
       \\[0.2cm]
  \hline \hline
\end{tabular}
\end{center}

\end{table}
\par
%
%
\begin{table}[h!]

\caption{First excited state wavefunction of $V^{(m)}(x)$}

\begin{center}
\begin{tabular}{ll}
  \hline\hline\\[-0.2cm]
  Type & $\psi_1(x)$ \\[0.2cm]
  \hline\\[-0.2cm]
  RHO & $x^{L+1} f^{-\frac{1}{2}\left(L+m+\frac{7}{2} + \alpha^{-m-1} \sqrt{B_{4m+2}}\right)}$ \\[0.2cm]
  & $\times \left[-2L-3 + 2\sqrt{B_{4m+2}} \sum_{k=1}^{m+1} \binom{m+1}{k} \alpha^{k-m-1} x^{2k}
       \right]$ \\[0.2cm]
  & $\times \exp\left[- \frac{1}{2} \sqrt{B_{4m+2}} \sum_{l=1}^m \frac{1}{l} \binom{m}{l} \alpha^{l-m-1}
       x^{2l}\right]$ \\[0.2cm]
  KC & $x^{L+1} f^{mL+m-1 + m\alpha^{-m-1} \sqrt{B_{2m}}}$ \\[0.2cm]
  & $ \times \left[-2L-3 - (m+1)(2L+3) \alpha x + 2 \sqrt{B_{2m}} \sum_{k=2}^{m+1} \binom{m+1}{k}
       \alpha^{k-m-1} x^k\right]$ \\[0.2cm]
  & $\times \exp\left[- \sqrt{B_{2m}} \sum_{l=1}^m \frac{1}{l} \binom{m}{l} \alpha^{l-m-1} x^l \right]$ 
       \\[0.2cm]
  M & $f^{\frac{m}{2} -1 + m \frac{\Delta}{2\alpha} + m \alpha^m \sqrt{B_{2m}}}$ \\[0.2cm]
  & $\times \left[-2\alpha-\Delta - (m+1) \left(\frac{\Delta}{\alpha}+2\right) e^x + 2 \sqrt{B_{2m}}
       \sum_{k=2}^{m+1} \binom{m+1}{k} \alpha^{m+1-k} e^{kx}\right]$ \\[0.2cm] 
  & $ \times \exp\left\{\left[\frac{1}{2} \left((m+1)\frac{\Delta}{\alpha} + m\right) + m \alpha^m 
       \sqrt{B_{2m}}\right] x\right\}$ \\[0.2cm]
  & $\times \exp\left[- \sqrt{B_{2m}} \sum_{l=1}^m \frac{1}{l} \binom{m}{l} \alpha^{m-l} e^{lx}\right]$ 
       \\[0.2cm]
  \hline \hline
\end{tabular}
\end{center}

\end{table}
\par
%
%
In Figs.~3 to 8, some examples of extended potentials $V^{(m)}(x)$, defined in Tables 3 and 5, are plotted, as well as their corresponding (unnormalized) wavefunctions $\psi_0(x)$ and $\psi_1(x)$.\par
%
%
\section{Conclusion}

In the present paper, we have shown that it is possible to generate infinite families of PDM Schr\"odinger equations with known ground and first excited states in DSUSY by considering extensions of known potentials endowed with a DSI property. For such purpose, we have combined two different approaches.\par
%
%
The first one is a generating function method, which enables to construct the first two superpotentials of a DSUSY hierarchy, as well as the first two partner potentials and the first two eigenstates of the first potential, from some generating function $W_+(x)$ [and its accompanying function $W_-(x)$].\par
%
%
The second approach is the CDSI method, wherein the DSI property of the starting potential is generalized to their extensions by adding some constraints on the parameters and by imposing compatibility conditions between the sets of constraints associated with the first two DSUSY steps. It is worth observing that in the constant mass case and conventional SUSY, the corresponding CSI method has encountered some problems for the extended radial potentials considered in Refs.~\cite{chakra, bera}, because the compatibility conditions lead to unphysical values of the angular momentum. In contrast, for the same type of extensions, the PDM background considered here, which provides an additional parameter, leads to more flexibility and removes such a type of difficulty.\par
%
%
In this work, we have given detailed results for some extensions of the linear and radial harmonic oscillators, as well as the Kepler-Coulomb and Morse potentials. Applying the methods presented here to other pairs of mass and potential with a DSI property would be an interesting topic for future investigation.\par
%
%
Taking into account the large number of physical PDM problems wherein the oscillator potential provides a first crude approximation and some anharmonic terms are needed to reproduce experimental data, it is obvious that the exact results presented here for potentials including anharmonic effects may find some interesting applications to practical problems. The search for such applications (as well as those for the extended Kepler-Coulomb and Morse potentials) would be another interesting topic for future investigation.\par
%
%
\section*{Acknowledgments}

The author is deeply grateful to some anonymous referees for very helpful suggestions and to Prof.\ V.M.\ Tkachuk for communicating Ref.~\cite{voznyak}.\par
%
%

%
%
\newpage

\begin{figure}
\begin{center}
\includegraphics{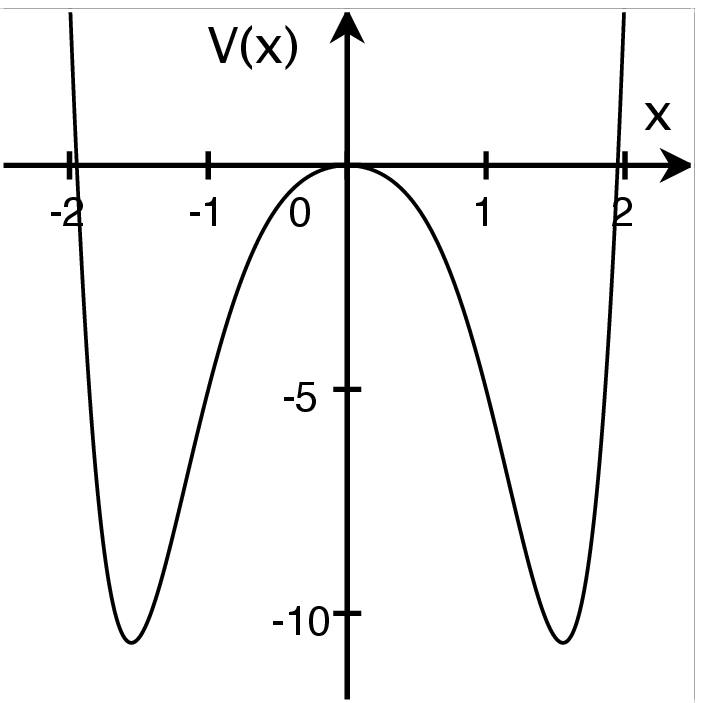}
\caption{Plot of  extended HO potential with $m = \alpha = B_6 = 1$. The ground and first excited state energies are $E_0 = 0$, $E_1 = 3$.}
\end{center}
\end{figure}
\par
%
%
\begin{figure}
\begin{center}
\includegraphics{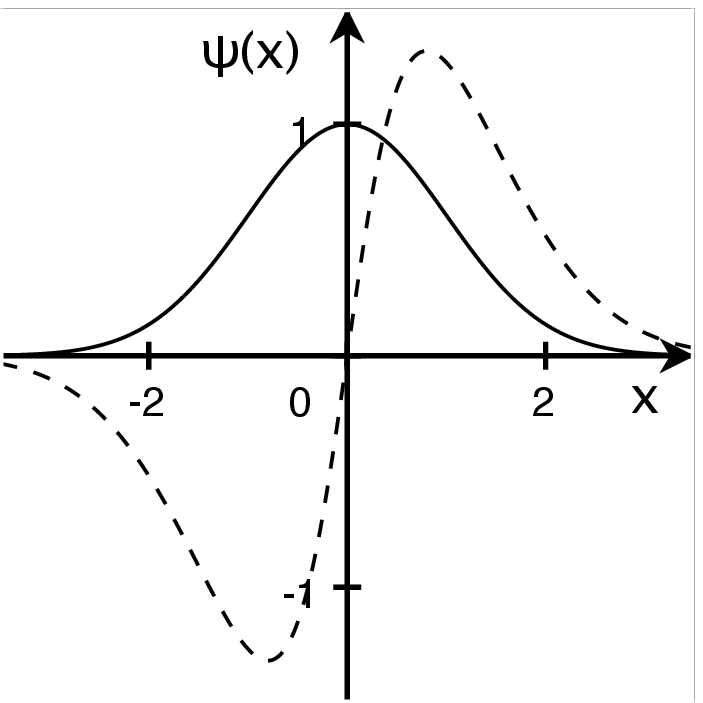}
\caption{Plots of the ground state wavefunction $\psi_0(x)$ (solid line) and of the first excited state wavefunction $\psi_1(x)$ (dashed line) for the potential displayed in Fig.~1.}
\end{center}
\end{figure}
\par
%
%
\begin{figure}
\begin{center}
\includegraphics{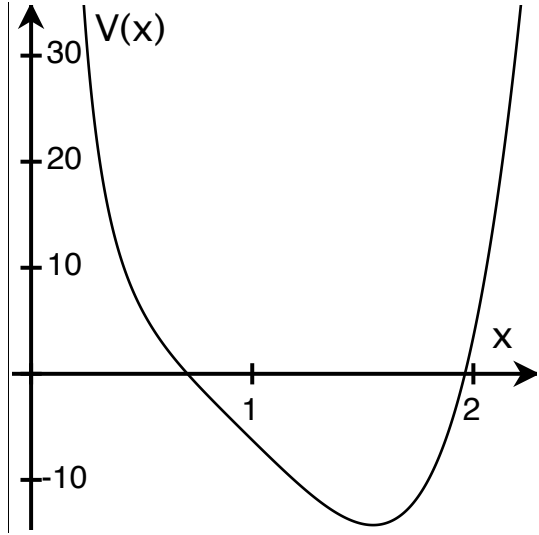}
\caption{Plot of extended RHO potential with $m = \alpha = L = B_6 = 1$. The ground and first excited state energies are $E_0 = 9/2$ and $E_1 = 65/2$.}
\end{center}
\end{figure}
\par
%
%
\begin{figure}
\begin{center}
\includegraphics{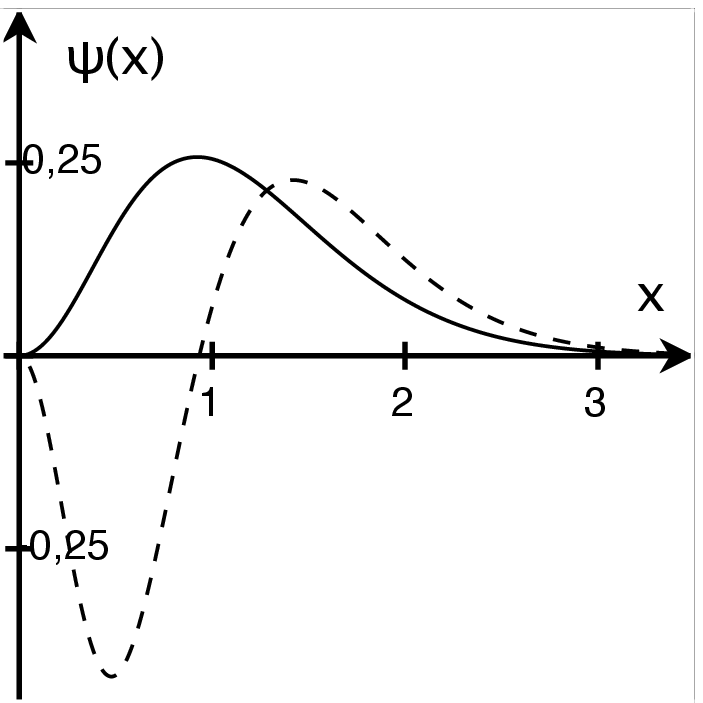}
\caption{Plots of the ground state wavefunction $\psi_0(x)$ (solid line) and of the first excited state wavefunction $\psi_1(x)$ (dashed line) for the potential displayed in Fig.~3.}
\end{center}
\end{figure}
\par
%
%
\begin{figure}
\begin{center}
\includegraphics{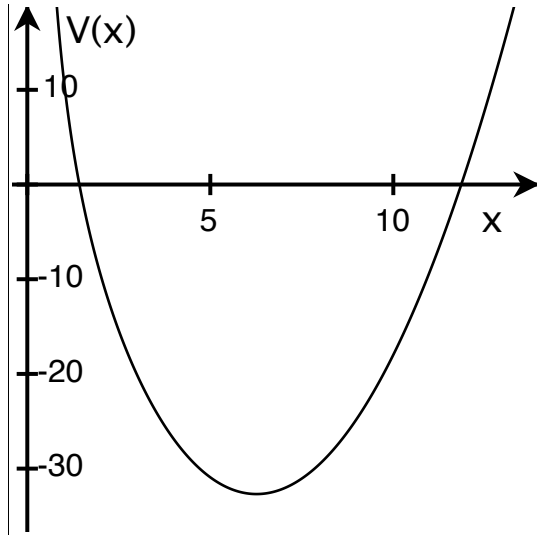}
\caption{Plot of extended KC potential with $m = \alpha = L = B_2 = 1$. The ground and first excited state energies are $E_0 = -99/4$ and $E_1 = -45/4$.}
\end{center}
\end{figure}
\par
%
%
\begin{figure}
\begin{center}
\includegraphics{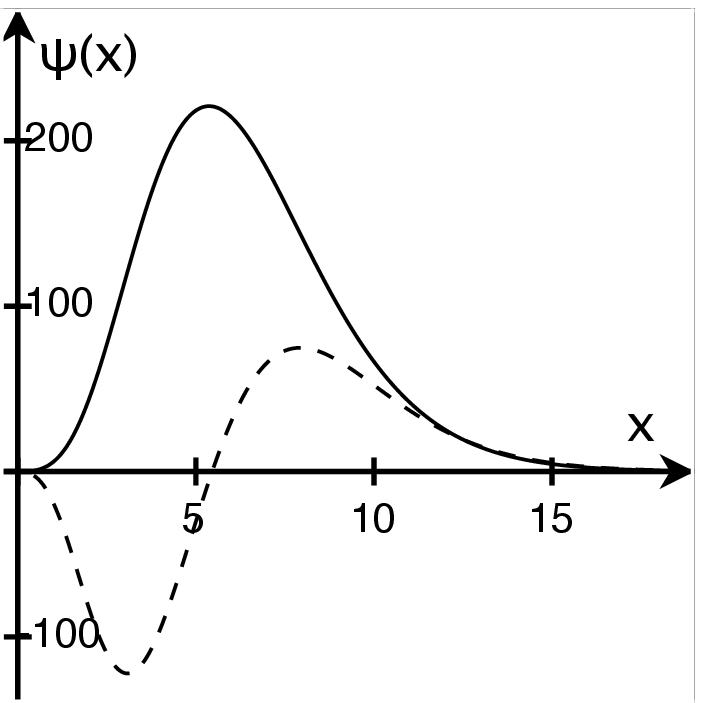}
\caption{Plots of the ground state wavefunction $\psi_0(x)$ (solid line) and of the first excited state wavefunction $\psi_1(x)$ (dashed line) for the potential displayed in Fig.~5.}
\end{center}
\end{figure}
\par
%
%
\begin{figure}
\begin{center}
\includegraphics{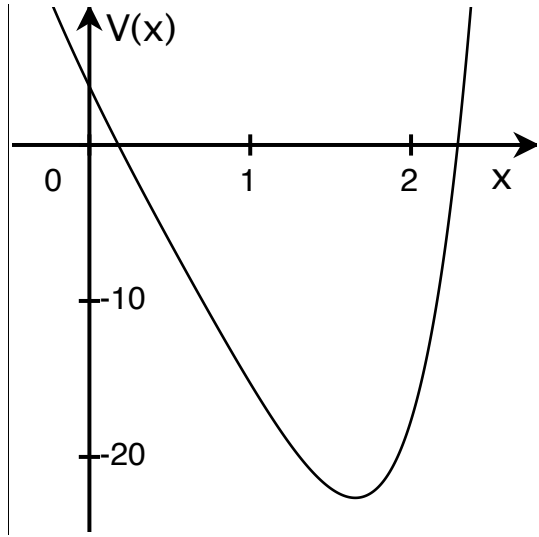}
\caption{Plot of extended M potential with $m = \alpha = B_2 = 1$ and $B_{-2} = 3/4$. The ground and first excited state energies are $E_0 = -65/4$ and $E_1 = -17/4$.}
\end{center}
\end{figure}
\par
%
%
\begin{figure}
\begin{center}
\includegraphics{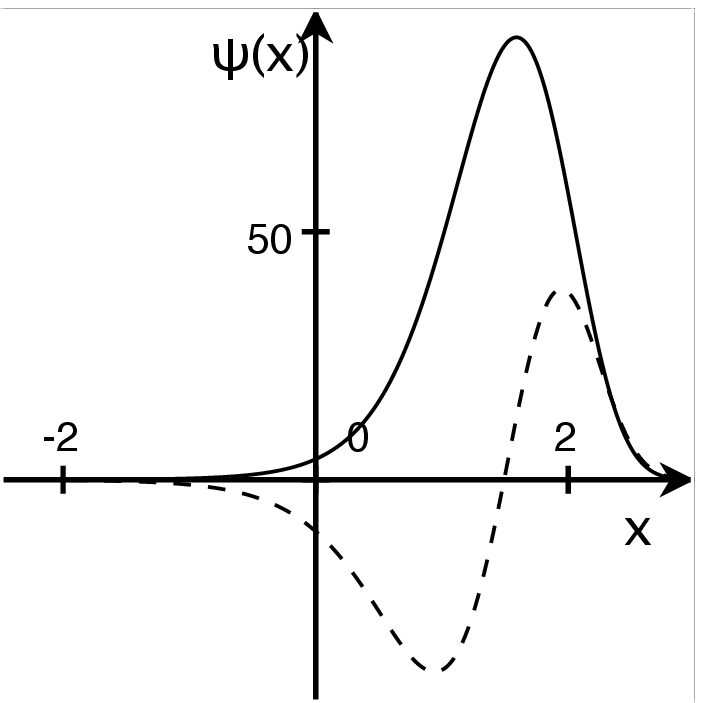}
\caption{Plots of the ground state wavefunction $\psi_0(x)$ (solid line) and of the first excited state wavefunction $\psi_1(x)$ (dashed line) for the potential displayed in Fig.~7.}
\end{center}
\end{figure}
\par

\end{document}